\documentclass[prb,onecolumn,floatfix,12pt,notitlepage,a4paper]{revtex4-1}
\usepackage{graphicx}
\usepackage{lineno}

\newlength{\onecolfig}
\newlength{\twocolfig}
\newcommand{\ion}[2]{\mbox{$^{#2}$#1$^+$}}
\newcommand{\Ca}[1]{\ion{Ca}{#1}}

\newcommand{\hfslev}[3]{\mbox{#1$^{\mbox{\tiny$#3$}}_{\mbox{\tiny$#2$}}$}}

\newcommand{\unit}[1]{\,\mbox{#1}}

\newcommand{\kHz}{\unit{kHz}}
\newcommand{\MHz}{\unit{MHz}}
\newcommand{\GHz}{\unit{GHz}}
\newcommand{\THz}{\unit{THz}}

\newcommand{\mW}{\unit{mW}}

\newcommand{\um}{\unit{$\mu$m}}
\newcommand{\nm}{\unit{nm}}

\newcommand{\us}{\unit{$\mu$s}}
\newcommand{\ns}{\unit{ns}}

\newcommand{\degree}{\mbox{$^{\circ}$}}

\newcommand{\mT}{\unit{mT}}

\newcommand{\ish}{\mbox{$\sim$}\,}
\newcommand{\ltish}{\protect\raisebox{-0.4ex}{$\,\stackrel{<}{\scriptstyle\sim}\,$}}

\newcommand{\up}{\mbox{$\uparrow$}}
\newcommand{\down}{\mbox{$\downarrow$}}

\newcommand{\ket}[1]{\mbox{$\left| #1 \right>$}}

\newcommand{\qr}{\marginpar{}}   

\begin{document}

\title{Fast quantum logic gates with trapped-ion qubits}

\author{V.~M.~Sch\"{a}fer, C.~J.~Ballance, K.~Thirumalai, L.~J.~Stephenson, T.~G.~Ballance, A.~M.~Steane and D.~M.~Lucas}

\affiliation{Department of Physics, University of Oxford, Clarendon Laboratory, Parks Road, Oxford OX1 3PU, U.K.}

\date{14 Nov 2017}

\maketitle

{\bf
Quantum bits based on individual trapped atomic ions constitute a promising technology for building a quantum computer~\cite{Wineland1998}, with all the elementary operations having been achieved with the necessary precision for some error-correction schemes~\cite{Harty2014,Ballance2016,Gaebler2016}\qr. However, the essential two-qubit logic gate used for generating quantum entanglement has hitherto always been performed in an adiabatic regime, where the gate is slow compared with the characteristic motional frequencies of ions in the trap~\cite{Ballance2016,Gaebler2016,Turchette1998,Leibfried2003,Benhelm2008}, giving logic speeds of order 10\kHz. There have been numerous proposals for performing gates faster than this natural ``speed limit'' of the trap~\cite{GarciaRipoll2003,Duan2004,GarciaRipoll2005,Steane2014,Palmero2017}. We implement the method of Steane {\em et al.}~\cite{Steane2014}, which uses tailored laser pulses: these are shaped on 10\ns\ timescales to drive the ions' motion along trajectories designed such that the gate operation is insensitive to optical phase fluctuations\qr. This permits fast (MHz-rate) quantum logic which is robust to this important source of experimental error. We demonstrate entanglement generation for gate times as short as 480\ns; this is less than a single oscillation period of an ion in the trap, and 8 orders of magnitude shorter than the memory coherence time measured in similar calcium-43 hyperfine qubits. The method's power is most evident at intermediate timescales, where it yields a gate error more than ten times lower than conventional techniques; for example, we achieve a 1.6\us\ gate with fidelity 99.8\%. Still faster gates are possible at the price of higher laser intensity. The method requires only a single amplitude-shaped pulse and one pair of beams derived from a continuous-wave laser, and offers the prospect of combining the unrivalled coherence properties~\cite{Bollinger1991,Harty2014,Wang2017}, operation fidelities~\cite{Harty2014,Ballance2016,Gaebler2016} and optical connectivity~\cite{Moehring2007} of trapped-ion qubits with the sub-microsecond logic speeds usually associated with solid state devices~\cite{Barends2014,Veldhorst2015}.}

Deterministic entanglement of multiple qubits, an essential pre-requisite for general quantum information processing, was first achieved nearly twenty years ago using laser manipulation of qubits stored in the hyperfine ground states of trapped atomic ions~\cite{Turchette1998}. Since then technical progress, the development of more robust methods, and improved understanding of error sources have yielded a steady improvement in the precision of the fundamental two-qubit quantum logic gate, with the gate error $\epsilon_g$ falling by approximately a factor of two every two years, to reach the level $\epsilon_g\approx 0.1\%$ in recent experiments~\cite{Ballance2016,Gaebler2016}. All elementary single-qubit operations have also been demonstrated with errors $<0.1\%$~\cite{Harty2014,Ballance2016,Gaebler2016}. These error levels are already an order of magnitude below the threshold level required for fault-tolerant quantum error correction schemes~\cite{Fowler2012}. In contrast the two-qubit gate speed has remained fairly constant since the first demonstrations; the gates with the lowest reported errors had durations of 30\us\ and 100\us. For qubits based on solid state platforms, the interactions are much stronger, allowing significantly faster two-qubit operations (typically \ish 50\ns\ for superconducting circuits~\cite{Barends2014}, and 480\ns\ for the recently-demonstrated gate in silicon-based qubits~\cite{Veldhorst2015}), but also leading to much shorter qubit coherence times (typically $T_2^* \ish 100\us$, compared with $T_2^* \ish 1$ minute for atomic systems). Substantial progress has also been made in demonstrating simple algorithms and quantum simulations involving \ish 10 qubits, and in developing technologies amenable to scaling to larger numbers of qubits~\cite{Monroe2013,Devoret2013}.

\begin{figure}
\includegraphics[width=0.7\twocolfig]{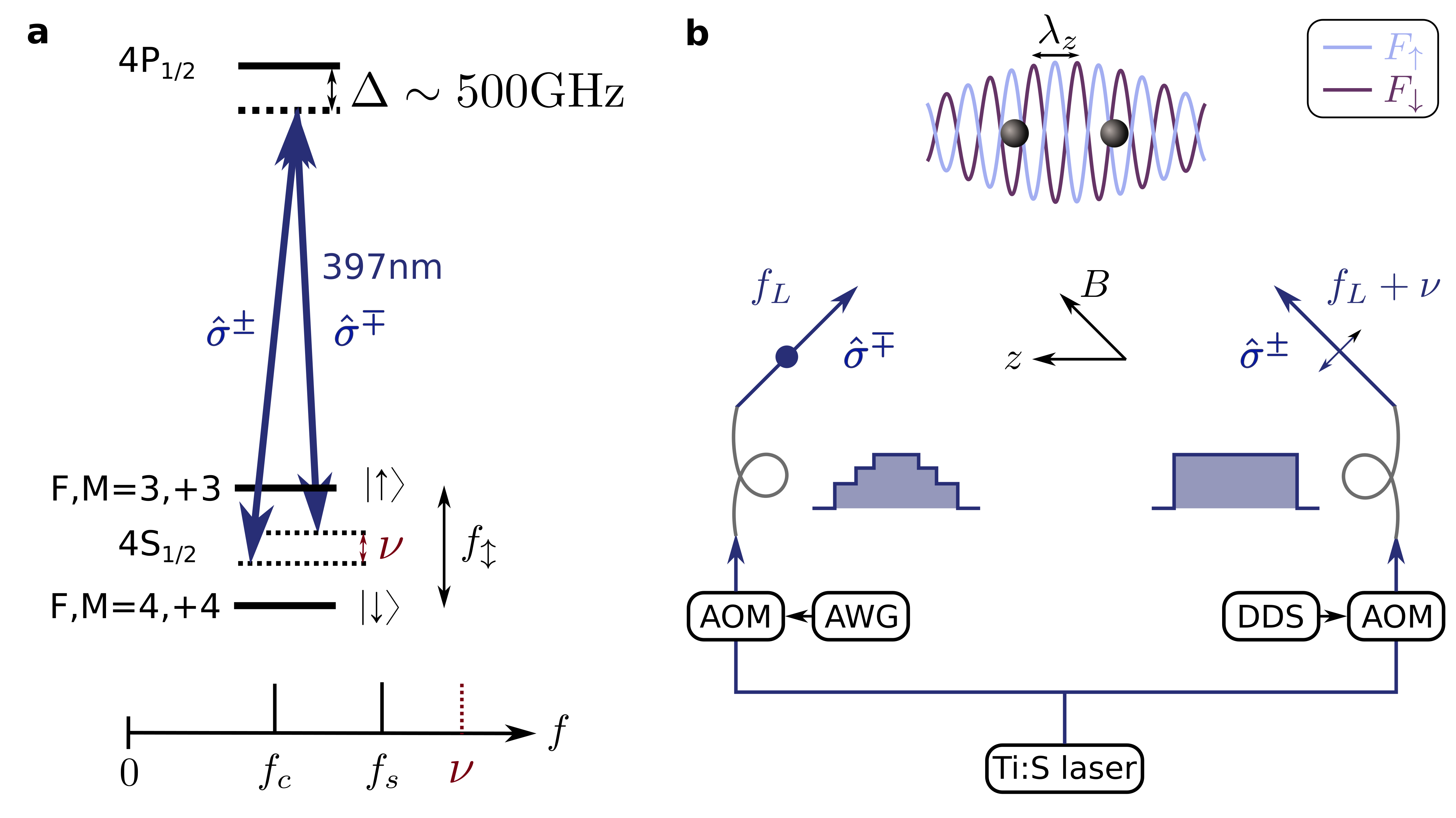}
\caption{%
Qubit states and Raman beam geometry.
(a) Qubits are stored in \Ca{43} hyperfine states \ket{\downarrow}=\hfslev{4S}{1/2}{4,+4} and \ket{\uparrow}=\hfslev{4S}{1/2}{3,+3}, with separation $f_\updownarrow=2.87\GHz$. Ion axial motional frequencies are $(f_c,f_s=1.92,3.33\MHz)$. The Raman beam difference frequency $\nu=3.43f_c$ for the fastest gate, while $f_c < \nu < f_s$ for the highest-fidelity gates.  
(b) One Raman beam propagates parallel to the quantization axis, set by a magnetic field $B\approx 14.6\mT$. The beams are perpendicular, such that their difference $k$-vector is parallel to the trap axis $z$, and have $\approx35\um$ waists at the ions, powers of up to 200\mW, and orthogonal linear polarizations. Their interference creates a polarization ``travelling standing wave'' (period $\lambda_z\approx 397\nm/\sqrt{2}$) that induces a spin-dependent force $F_\downarrow, F_\uparrow$ on the ions. High-bandwidth acousto-optic modulators (AOMs) shape the laser pulses on \ish 10\ns\ timescales; we use a constant-amplitude pulse for one beam, and an amplitude-shaped pulse for the other beam.
}
\label{F:beams}
\end{figure}

In previous trapped-ion work the speed of the two-qubit gate operation has been limited by the use of methods that operate in an adiabatic regime with respect to the secular motional frequencies of the ions; as these are typically \ish1\MHz, gate durations are generally $\gg 1\us$, and attempts to increase the gate speed have resulted in larger gate errors (for example, $\epsilon_g=3\%$ at the shortest reported gate time of $t_g=5.3\us$\cite{footnote1}). With recent progress in demonstrating faster techniques of ground-state laser cooling~\cite{Lin2013}, ion shuttling~\cite{Bowler2012,Ruster2014}, and qubit readout~\cite{Noek2013b}, present two-qubit gate speeds threaten to be the limiting factor in the clock speed of a trapped-ion processor based on a ``quantum CCD'' architecture~\cite{Wineland1998}, especially given that error-correction circuits typically contain more gates than state preparation and readout operations. The two-dimensional QCCD architecture would be a natural choice for implementing surface-code error correction methods~\cite{Fowler2012}, although these can also be mapped onto one-dimensional ion chains\qr. Errors due to ambient heating of the ions' motion are proportional to $t_g$ and will thus be suppressed for fast gates, which is advantageous for microfabricated traps where the ions are confined near to electrode surfaces and hence subject to greater electric field noise~\cite{Turchette2000}. Spin-dephasing errors due to, e.g., magnetic field fluctuations (which typically have a $1/f$ noise spectrum), will likewise be reduced, allowing the use of qubit states which have first-order sensitivity to magnetic field~\cite{Ruster2016} (at least during gate operations, as here). 

The ``speed limit'' set by the trap frequency $f_c$ is not a fundamental barrier: the Coulomb interaction responsible for coupling the ions is almost instantaneous at the typical separation of trapped ions (3.5\um\ in our work), and there have been a variety of theoretical proposals for fast gates with $t_g \ltish 1/f_c$, for example
refs.~\cite{GarciaRipoll2003,Duan2004,GarciaRipoll2005,Steane2014,Palmero2017}. None of these has so far been demonstrated~\cite{footnote2}\qr. Here, after first exploring the limits of the conventional $\sigma_z\otimes\sigma_z$ gate mechanism originally demonstrated by Leibfried {\em et al.}~\cite{Leibfried2003}, we implement the scheme proposed by Steane {\em et al.}~\cite{Steane2014}, in which the single rectangular laser pulse used in the conventional adiabatic method is replaced by a pulse whose amplitude is shaped in time.

The operation of the gate relies on a qubit-state-dependent force, which originates from the spatially-varying light shift caused by a ``travelling standing wave'', generated by the optical interference pattern of two non-copropagating laser beams with difference frequency $\nu$ (fig.~\ref{F:beams}). We specialize to the case of two ions with the force coupling only to the axial modes of motion. We discuss the behaviour in three regimes: (1) a single rectangular pulse in the adiabatic regime, (2) a single rectangular pulse in the non-adiabatic regime, (3) a fast shaped pulse or pulses. 

\begin{figure*}
\includegraphics[height=9.4cm]{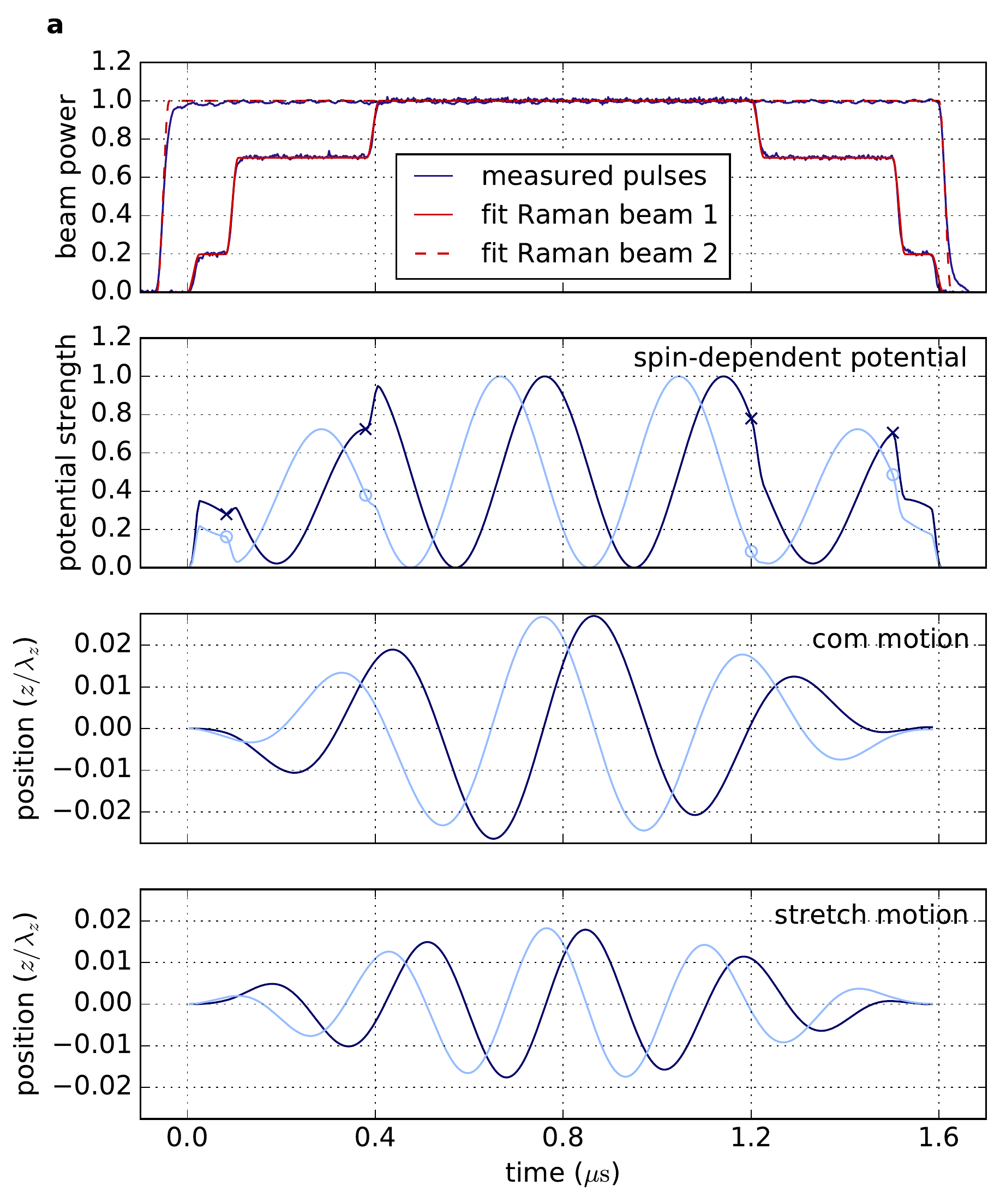}
\includegraphics[height=9.4cm]{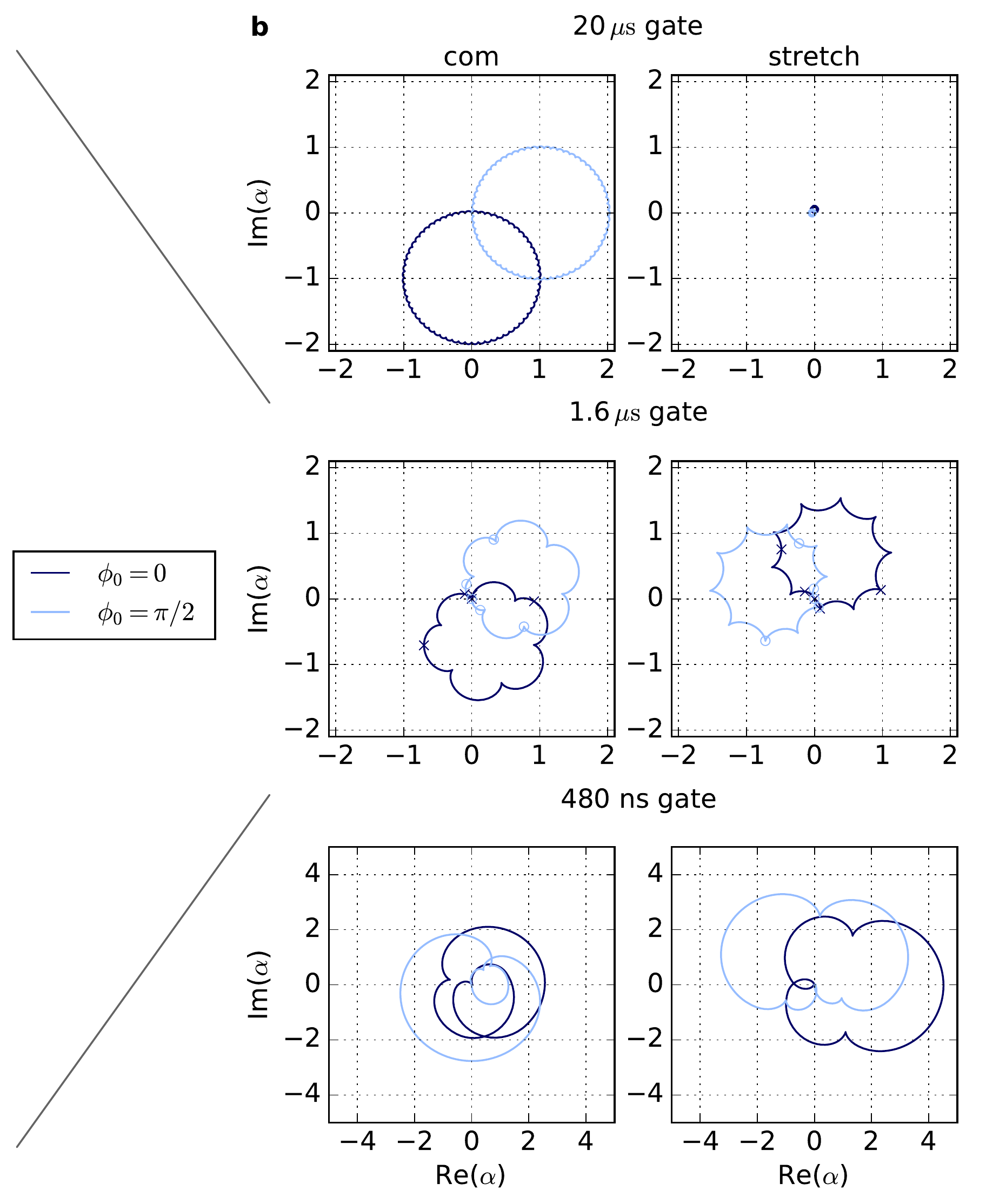}
\caption{%
Optical beat notes and motional trajectories of the ions, for two example initial optical phases $\phi_0=(0,\pi/2)$ rad.
(a) For the $t_g=1.6\us$ gate the plots show, from top: the Raman laser pulses; their (calculated) optical beat note, which gives rise to the spin- and position-dependent potential (and hence force) that the ions experience; the ions' centre-of-mass displacement; their stretch-mode  displacement. The beat frequency is $\nu=2.63\MHz\approx 1.37f_c$. The force and motions clearly depend on $\phi_0$; however, the pulse shape is designed such that, for all $\phi_0$, both trajectories return to zero displacement at $t=t_g$.
(b) Phase-space trajectories (rotating frame) for gates in three regimes. For a conventional adiabatic ($t_g=20\us$) gate, $\nu\approx 1.03f_c$ and the stretch mode is barely excited; $\phi_0$ affects the orientation of the (nearly-circular) trajectory, but not its shape or area. For $t_g=1.6\us$, both modes are driven and $\phi_0$ affects the shape of the trajectories slightly; amplitude shaping is necessary to close the loops for both modes and to ensure the net gate phase is independent of $\phi_0$. (Symbols correspond to steps in the pulse amplitude.) For $t_g=480\ns<1/f_c$ the trajectory depends strongly on $\phi_0$; this illustration makes the Lamb-Dicke approximation, but out-of-Lamb-Dicke effects mean that the loops no longer close, leading to significant gate errors.
}
\label{F:trajectories}
\end{figure*}

Case 1. By choosing $\nu = f_c + \delta$ with $\delta \ll f_c$,
only the centre-of-mass normal mode at frequency $f_c$ is excited (to first approximation) and the rotating wave approximation holds for the treatment of
the motion. Starting from a state cooled to the Lamb-Dicke regime ($\eta^2 n \ll 1$, where $\eta$ is the Lamb-Dicke parameter and $n$ the motional quantum number), the motion traces out an approximately circular path in the (rotating frame) phase space of the harmonic oscillator, returning to its starting point after time $t_g = 1/\delta$ (fig.~\ref{F:trajectories}). The geometric gate phase $\Phi$ is determined by the (signed) area enclosed by this path, which is proportional to $\Omega^2$, where $\Omega$ is the Rabi frequency. We require $\Phi=\pi/2$ to generate the maximally-entangled state $(\ket{\down\down}+{\mathrm i}\ket{\down\up}+{\mathrm i}\ket{\up\down}+\ket{\up\up})/2$ from the separable state $(\ket{\down\down}+\ket{\down\up}+\ket{\up\down}+\ket{\up\up})/2$ after time $t_g$. The gate phase $\Phi$ is independent of both the initial motional state (within the Lamb-Dicke regime), and the phase $\phi_0$ of the optical beat note at the start time $t=0$. The latter is crucial for achieving high gate fidelity in the laboratory, because $\phi_0$ is sensitive to nanometre-scale length differences between the two laser beam paths. Such gates were implemented previously~\cite{Leibfried2003, Ballance2016}.

\begin{figure*}
\includegraphics[width=\onecolfig]{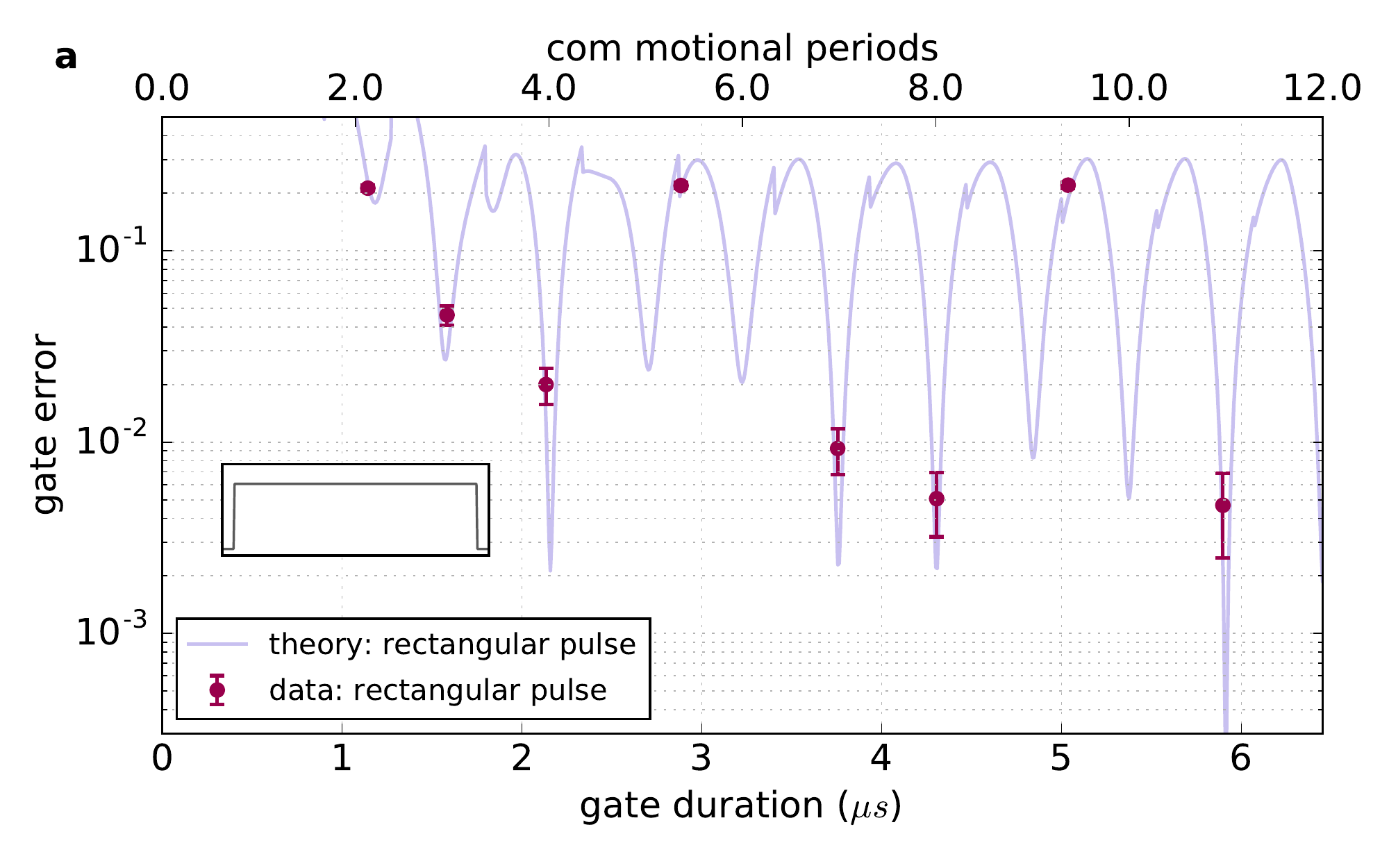}
\includegraphics[width=\onecolfig]{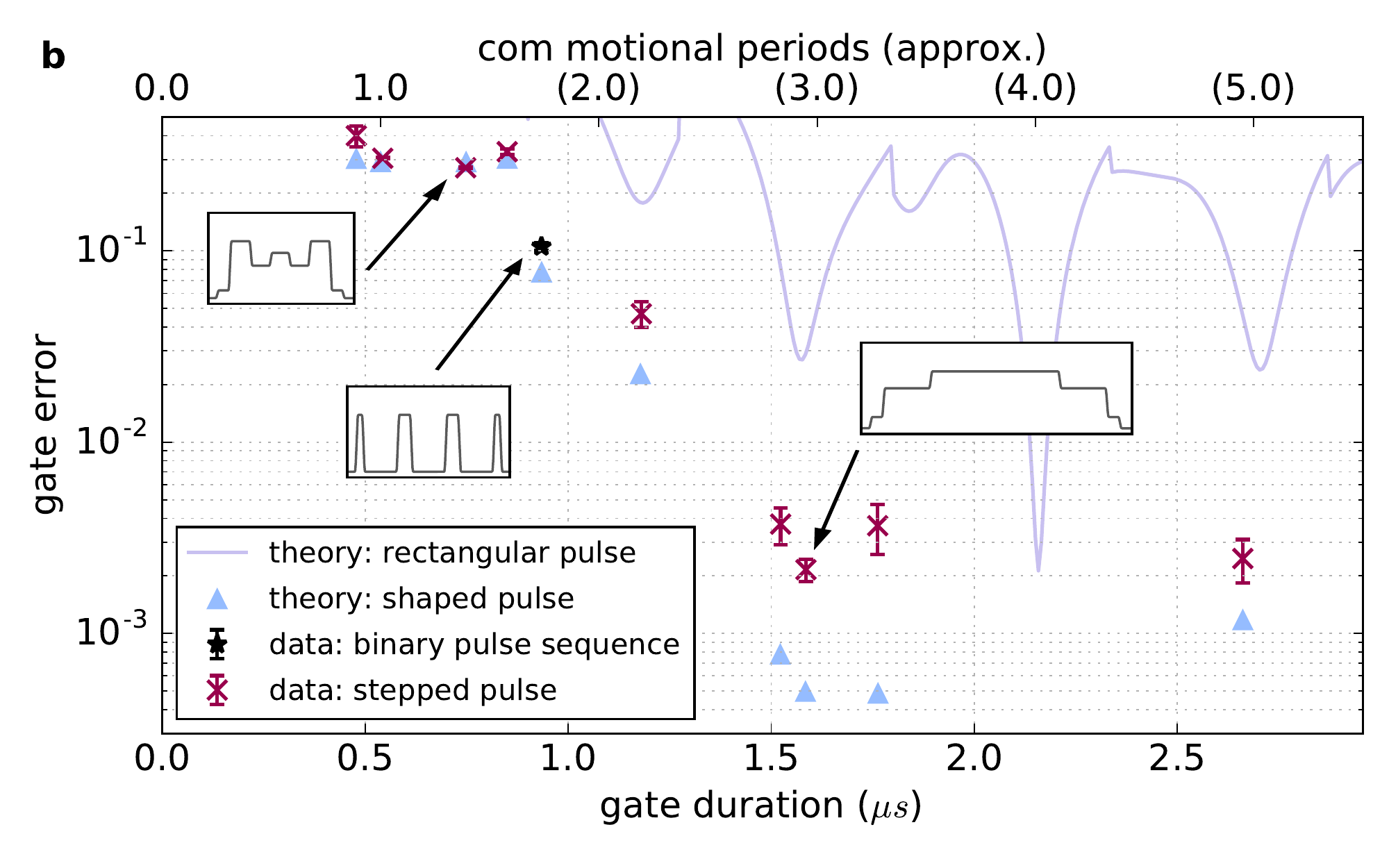}
\caption{%
Theoretical and experimental two-qubit gate errors (error bars give 1$\sigma$ statistical errors). 
(a) Conventional single rectangular pulse. The curve shows the coherent error achievable (i.e.\ excluding photon scattering and technical errors). At each time, $\Omega$ and $\nu$ are adjusted to minimize $\epsilon_g$; discontinuities occur where the optimum $\nu$ switches between $f_c<\nu<f_s$ and $\nu>f_s$. (Although significant reduction in gate error can be made by shaping the pulse edges~\cite{Ballance2016} when $t_g \gg 1/f_c$, negligible improvement is possible when the gate duration becomes comparable to the shaping time constant, at $t_g \ltish 4.5/f_c$.) Data points show experimentally-measured gate errors. 
(b) Amplitude-shaped pulses. The curve is repeated from (a) for comparison. Simulated errors (triangles) are dominated by out-of-Lamb-Dicke effects for $t_g<1.5\us$. The other points are measured gate errors, after optimizing pulse-shape parameters using real-time feedback from the experiment. Insets illustrate example pulse shapes. 
}
\label{F:errortime}
\end{figure*}

Case 2. The gate speed is increased by increasing $\delta$, but for $\delta \sim f_c$ there are three complicating factors: firstly, both the centre-of-mass mode and the stretch mode (at $f_s = f_c \sqrt{3}$) of a two-ion crystal will be excited and the associated trajectories in phase space will not in general close at the same time; secondly the trajectories depend on $\phi_0$; thirdly there is a time-dependent light shift independent of the motion but which also depends on $\phi_0$ and can result in a large single-qubit phase $\phi_{\rm LS}$. Consequently the expected gate error has a complicated dependence on gate time, and rises steeply as the gate time approaches the period of the motion. This is shown in fig.~\ref{F:errortime}a, together with a selection of results achieved in our experiments. We measure a gate error $\epsilon_g = 2.0(5)$\%
for $t_g = 2.13\,\mu$s, and the theory shows that no solutions exist with errors below this at shorter times.

Case 3. Replacing the single rectangular pulse of the conventional method by a shaped pulse gives more degrees of freedom (i.e.\ those parameters describing the pulse shape), which can be exploited to find especially well-performing or ``magic'' pulses. In particular, we want to achieve all of the following: that the phase space trajectories for both modes should close simultaneously at $t = t_g$; that the appropriate sum of (signed) areas enclosed be independent of $\phi_0$, even though the trajectories themselves may depend on $\phi_0$; that the light-shift induced phase $\phi_{\rm LS}$ be independent of $\phi_0$ and preferably small; that the pulse area is small to minimize photon scattering~\cite{Ozeri2007} and that the gate error be not too sensitive to errors in parameter settings. A shaped pulse or pulse-sequence is deemed a `solution' when it has all these properties, such that the gate error predicted for a perfectly realized sequence is below an upper bound $\epsilon_t$ set by practical considerations. That is, one sets $\epsilon_t$ well below the error one is prepared to accept in the laboratory, and seeks solutions by numerical search.

Several classes of solution are given by Steane {\em et al.} for particularly simple pulse shapes. We implemented two types of time-symmetric sequence: a binary pulse sequence (where a constant amplitude force is simply switched on and off), and five- or seven-segment ``stepped" pulses. Example phase-space trajectories are shown in fig.~\ref{F:trajectories}. By this means we obtained gates up to an order of magnitude faster than those previously demonstrated. However, to understand the experimentally observed gate error, and the optimal pulse shapes, we had to develop the theory further. 

The solutions given in ref.~\cite{Steane2014} assume that the motion remains within the Lamb-Dicke regime; for $t_g\ish 1/f_c$ this is a poor approximation, as large excursions in phase space are required to enclose sufficient area. For large excursions, the ions become sensitive to the spatial variation of the force, leading to modification of the trajectories and squeezing of the motional wavepackets~\cite{McDonnell2007}\qr. We extended the theory with numerical modelling to include the effects of motional excursion beyond the Lamb-Dicke regime, and found solutions which give the minimum gate error for times in the range $200\ns < t_g < 5.0\us$ (see Methods). The most efficient solutions, giving optimal use of the available laser power, are found when $f_c < \nu < f_s$, where both modes are excited such that the geometric phases from each mode add constructively ($\Phi=\Phi_c+\Phi_s$); conversely, when $\nu > f_s$, the phases subtract and more laser power is required to achieve $\Phi=\pi/2$ (in turn leading to higher photon scattering error~\cite{Ozeri2007}). The numerically-calculated errors for some of these efficient solutions are shown in fig.\ref{F:errortime}b, together with experimentally achieved gate errors for gate times between 480\ns\ and 2.7\us\ (see Methods for experimental details). The fastest gate time is slightly below the centre-of-mass motional period ($1/f_c=540\ns$), but the error is large (40\%). The binary pulse sequence achieves 11\%\ error at 0.93\us\ gate time. The minimum error measured is 0.22(3)\% at $t_g=1.6\us$, using a stepped pulse, which is close to the lowest two-qubit gate errors previously reported~\cite{Ballance2016,Gaebler2016}, whilst being 20--60 times faster. This error is an order of magnitude lower than that achievable with the conventional single-pulse method at the same $t_g$. For the 1.6\us\ gate, we estimate the total error due to known sources to be $\approx 0.18\%$ (Table~\ref{T:error_sources}).

\begin{table}
\centering
\begin{tabular}{|l|r|r|}
\hline
error source & \multicolumn{1}{c|}{$t_g=1.6\us$} & \multicolumn{1}{c|}{$t_g=480\ns$}\\
\hline
out-of-Lamb-Dicke effects & $5\times 10^{-4}$& $3\times 10^{-1}$ \\
optical phase chirp & $\sim 4\times 10^{-4}$ & $\sim 6\times 10^{-3}$ \\
pulse timing and amplitudes & $\sim 2\times 10^{-4}$ & $\sim 1\times 10^{-3}$ \\
radial mode excitation & $\ltish4\times 10^{-5}$ & $\ltish4\times 10^{-3}$\\
photon scattering & $6\times 10^{-4}$ & $7\times 10^{-3}$ \\
centre-of-mass heating rate & $8\times 10^{-5}$ & $3\times 10^{-5}$ \\
\hline 
total error & $1.8\times 10^{-3}$ & $3.3\times 10^{-1}$ \\
\hline
\end{tabular}
\caption{Error budget for the highest-fidelity and fastest gates achieved. The total is the linear sum of the individual errors; this assumes they are constant and add incoherently.}
\label{T:error_sources}
\end{table}

In our setup the gate speed and fidelity are limited by the breakdown of the Lamb-Dicke approximation for $t_g\ltish 1/f_c$. Faster and/or higher fidelity gates are possible by reducing the Lamb-Dicke parameters (here $\eta_c=0.126, \eta_s=0.096$); for example, decreasing the $90\degree$ angle between the two laser beams (fig.~\ref{F:beams}b) to give $\eta_c=0.08$ would reduce the error contribution from out-of-Lamb-Dicke effects to $7\times 10^{-5}$\qr. This in turn requires higher laser intensities at the ions; although we use a moderately high laser power ($\ish$150\mW\ per beam for the fastest gate), the intensity is modest (\ish 0.1\unit{\mW/\um$^2$}) and the spot size ($w_0\approx 35\um$) could be significantly reduced. Alternatively, if the optical phase $\phi_0$ could be sufficiently well controlled, solutions can be found for fixed $\phi_0$ which allow faster gates and higher fidelities~\cite{Palmero2017}\qr.

In conclusion, we have demonstrated a fast (1.6\us), robust two-qubit gate method for trapped-ion qubits which combines state-of-the-art gate fidelity (99.8\%) with more than an order of magnitude increase in gate speed. At the fastest speed demonstrated (480\ns) the fidelity achieved (60\%) may not be useful for information processing, but might have other applications (such as quantum logic spectroscopy of short-lived exotic species~\cite{Schmidt2005,Meyer2000}; this would also require the use of fast laser cooling techniques~\cite{Machnes2010}). The method is technically simple, requiring only a single amplitude-shaped pulse from a cw laser, and the laser intensities required are within reach of miniature solid state violet diodes~\cite{Schafer2015}. These considerations are important if the techniques are ultimately to be scaled to the very large numbers of qubits necessary for an error-corrected quantum computer.

\section*{METHODS}

\subsection*{Numerical modelling}

\noindent Most trapped-ion experiments can be described in the Lamb-Dicke regime, i.e.\ the optical field is assumed to be uniform over the extent of each ion's wavefunction. However for the large phase-space displacements necessary to perform fast gates this assumption breaks down: the curvature of the field can no longer be neglected. This means the force experienced by an ion depends on its displacement in phase space and this leads to squeezing of the wavefunction, as well as modification of the motional trajectory.\qr

To model the coherent error of a given gate sequence, we therefore numerically integrate the full Hamiltonian (that is, without making the Lamb-Dicke approximation) using the split-operator method, explicitly averaging over different initial optical phases. As this is a computationally intensive process, the gate sequences used in the experiments were pre-selected by an efficient solver that works in the Lamb-Dicke regime. Following Steane {\em et al.}, we optimize candidate solutions starting from a random seed, and select a set of candidate solutions that have an error of $< 10^{-4}$ in the Lamb-Dicke approximation.

These candidate solutions were then evaluated using the full solver, and the most promising were optimized further. For the experiments, we chose solutions from this set by looking for a combination of low coherent error and low integrated pulse area (this both selects for a low photon scattering error~\cite{Ozeri2007}, and avoids fragile sequences that use large motional excitations, which are more sensitive to parameter variations).

Several different pulse shapes were evaluated. The seven-segment symmetric pulse shape offered a sufficient number of parameters to find a dense set of good solutions, whilst being easy to implement and to verify. The exact shape of the rising and falling edges is unimportant: the rise-time can be varied from zero to the segment length without a change in gate fidelity, providing that an overall scaling factor is applied to the gate Rabi frequency to compensate for the changing spectral content.

\subsection*{Raman beams}

\noindent The light source for the Raman beams is a frequency-doubled Ti:sapphire laser with 1.8W output power at $397\nm$~\cite{SolsTiS}. The Raman detuning was $\Delta=-1\THz$ for single rectangular pulses; for shaped-pulse gates with $t_g \leq1\us$, $\Delta=-200\GHz$ and for $t_g>1\us$, $\Delta=-800\GHz$. The detuning was changed to reduce photon scattering errors for gates requiring lower Rabi frequencies. For the fastest gate, peak powers of $192\mW$ and $96\mW$ were used for the two Raman beams, which had waists at the ions of 33\um\ and 38\um\ respectively ($1/e^2$ intensity radius). The ratios of Raman beam powers were chosen such that the scattering error was approximately minimized. The beams were modulated by a pair of acousto-optic modulators~\cite{AOM} (AOMs) with $24\ns$ rise time (10\%--90\%) to create the shaped pulses driving the gate. The amplitude-shaped radiofrequency (RF, $200\MHz$) signal for the stepped pulse was defined with an arbitrary waveform generator~\cite{AWG} (AWG) and fed to the first AOM. The second AOM was driven by a direct digital synthesis (DDS) source~\cite{DDS} (fig.~\ref{F:beams}b).

Phase chirps of the modulated beam were measured in an optical homodyne experiment and found to be significant during switching of the RF amplitude. 
Driving the AOM at its centre frequency ($200\MHz$) minimized the phase chirps~\cite{Degenhadt2005} such that their contribution to the gate error was small (table~\ref{T:error_sources}).

\subsection*{Pulse calibration}

\noindent Performing fast gates with high fidelities requires precise control of the pulse parameters. Due to the non-linear response of the AOM for different RF drive amplitudes, the pulse shape was measured on a photodiode and the relative drive amplitudes of each pulse segment were adjusted to match the measured amplitudes to their theoretically-predicted optimum levels. The relative amplitudes of the stepped pulses were set with $\pm 0.2\%$ accuracy. The waveform programmed into the 1.25\unit{Gsps} AWG had a 5.0\ns\ risetime in order to spread the pulse edges over several time points and improve the effective timing resolution\qr. The timing precision of the optical pulses was measured to be $0.2\ns$ (standard deviation of fitted pulse lengths). Setting the pulse-shape parameters to their theoretically-predicted values yielded optimal fidelity for all gate sequences where the Lamb-Dicke approximation held well. There are three remaining parameters characterising the gate sequence: the peak beam power, the Raman beat note frequency $\nu$ and the phase offset $\phi_{\pi/2}$ of the last $\pi/2$-pulse of the Ramsey interferometer ($\phi_{\pi/2}$ compensates for the single-qubit phase acquired during the gate). The beat note frequency and beam power were set to their theoretically-predicted values and then optimized empirically; in all cases the optimized values agreed well with their theoretical predictions. The peak pulse powers of each beam were stabilized at the beginning of each experimental sequence. The phase offset $\phi_{\pi/2}$ was calibrated empirically. Initially gate parameters were optimized with a Nelder-Mead algorithm. After the minimisation of optical phase chirps this was no longer necessary and linear optimisation of single parameters was found to be sufficient. A list of parameters for the fastest and highest-fidelity gates is given in table~\ref{T:gateparams}.

\begin{table}
\centering
\includegraphics[width=\onecolfig]{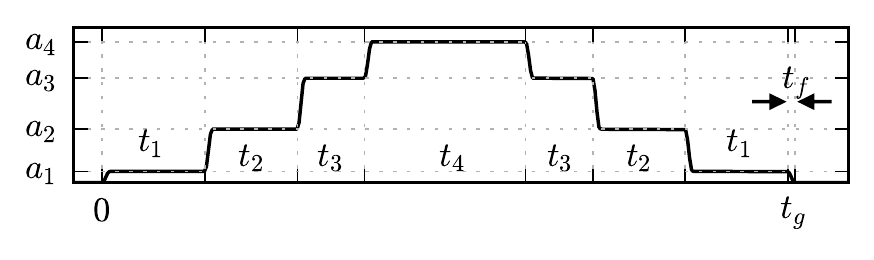}\\ 
\begin{tabular}{|@{~~~}l@{~~~}|@{~~~}r@{~~~}|@{~~~}r@{~~~}|}
\hline
{} & \multicolumn{2}{c|}{gate duration} \\ 
parameter & $t_g=483\ns$ & $t_g=1.59\us$\\
\hline
Raman detuning $\Delta$ & $-200\GHz$& $-800\GHz$\\
Raman beat note frequency $\nu$ & $6.3802\MHz$& $2.6301\MHz$\\
axial centre-of-mass frequency $f_c$ & $1.8615\MHz$ & $ 1.9243\MHz$ \\
peak power (pulse-shaped beam) & $192\mW$& $58\mW$\\
power (non-shaped beam)  & $96\mW$ & $48\mW$\\
single-qubit phase $\phi_{\pi/2}$& $91.4^\circ$ & $21.4^\circ$\\
pulse time $t_{1}$& $71.4\ns$& $82.1\ns$\\
pulse time $t_{2}$& $64.5\ns$& $299.9\ns$\\
pulse time $t_{3}$& $46.7\ns$& --- \\
pulse time $t_{4}$& $112.3\ns$ & $819.5\ns$ \\
pulse fall-time $t_f$& $5.0\ns$ & 5.0\ns \\
pulse amplitude $a_{1}$& $ 0.284$& $ 0.445$\\
pulse amplitude $a_{2}$& $ 0.617$& $ 0.838$\\
pulse amplitude $a_{3}$& $ 0.862$& --- \\
pulse amplitude $a_{4}$& $ 1$ & $ 1$ \\ \hline
\end{tabular}
\caption{(EXTENDED DATA) Gate parameters used for the fastest gate (7 segments) and for the highest-fidelity gate (5 segments). The pulse envelope above illustrates the definition of the pulse timing and amplitude parameters. The timing parameters refer to the timing of the waveform programmed into the AWG, for which a $t_f=5.0\ns$ rise/fall-time (0\%--100\%) was used; the measured rise/fall-time (10\%--90\%) of the laser pulses was 24\ns, due to the bandwidth of the particular AOMs used (see fig.\ref{F:trajectories}a). The waists ($1/e^2$ intensity radii) of the Raman beams were 33\um\ and 38\um\ for the pulse-shaped and non-shaped beams respectively.}
\label{T:gateparams}
\end{table}

\subsection*{Experimental procedure}

\noindent All gates were performed in a blade-type linear Paul trap~\cite{PhD:Gulde,MSc:Woodrow}, with axial centre-of-mass frequency $f_c=1.92\MHz$ for $t_g >1\us$, and $f_c=1.86\MHz$ for $t_g  \leq1\us$ and for all single rectangular-pulse gates. The axial frequency was changed after re-aligning the Raman beams to suppress coupling to radial modes. In both cases the axial frequency was chosen such that the ion spacing was $12\frac{1}{2}\lambda_z$, where $\lambda_z=283\nm$ is the periodicity of the travelling standing wave providing the gate force. 
The gate was performed on the qubit states $\ket{\downarrow}=\mathrm{4S}_{1/2}\ket{F=4,M=+4}$ and $\ket{\uparrow}=\mathrm{4S}_{1/2}\ket{F=3,M=+3}$ in $^{43}\mathrm{Ca}^+$ at $B=14.6\mT$. (This value of the $B$-field gives access to the ``atomic clock'' qubit $\ket{\downarrow'}=\mathrm{4S}_{1/2}\ket{F=4,M=0}$ and $\ket{\uparrow'}=\mathrm{4S}_{1/2}\ket{F=3,M=+1}$ with long coherence time, measured to be $T_2^*\ish 1\unit{minute}$ in prior work~\cite{Harty2014}, ideal for use as a memory qubit.) The ions were laser-cooled with dark-resonance Doppler cooling~\cite{Allcock2016} to $\bar{n}\approx1.8$ and further cooled with sideband cooling to $\bar{n}\lesssim 0.05$. After state preparation in $\ket{\down\down}$ we created an entangled state by placing the geometric phase-gate in one arm of a Ramsey interferometer split by a spin-echo $\pi$-pulse~\cite{Leibfried2003}. The gate errors were determined by using partial tomography~\cite{Sackett2000} to measure the fidelity of the created state with respect to the desired state $\left(\ket{\down\down}+\ket{\up\up}\right)/\sqrt{2}$.

\subsection*{Error analysis}

\noindent All gate errors and fidelities quoted are after correction for state-preparation and readout errors~\cite{Ballance2016}. The total state-preparation and readout error with two ions was typically $\bar{\epsilon}_{\mathrm{SPAM}}=1.4(1)\times 10^{-3}$ per ion (averaged over both qubit states).
The 0.22(3)\% error reported for the $t_g=1.6\us$ gate is the average of five experimental runs measured on two days. Directly after calibrating experimental parameters, the lowest error measured was 0.15(3)\%; an hour after calibration, the measured error was 0.28(3)\%. Quoted uncertainties are statistical only. We also measured the accumulated error for concatenated sequences of up to 7 gates, and found no evidence for coherent errors.\qr

Errors due to radial mode excitation are largest for gates around $t_g = 800\ns$, because here the Raman beat note frequency $\nu$ is close to resonance with the radial mode frequencies ($\approx 4.2\MHz$); with our final Raman laser beam alignment we can limit errors due to radial mode excitation to $\epsilon_g < 5\times10^{-2}$ at $t_g=800\ns$. An advantage of fast gates is that they are insensitive to errors associated with motional decoherence or heating; despite the relatively large heating rate of this trap ($\dot{\bar{n}}\approx100\,\mathrm{s}^{-1}$ for the axial centre-of-mass mode) the contribution to the gate error is negligible. 
A summary of the main errors present in our experiments, for the lowest-error gate, and for the fastest gate, is given in table~\ref{T:error_sources}.

\section*{Acknowledgements}
\noindent 
This work was supported by the U.K.\ EPSRC ``Networked Quantum Information Technologies'' Hub, and the U.K.\ Defence, Science and Technology Laboratory. VMS acknowledges funding from Balliol College, Oxford. CJB acknowledges funding from Magdalen College, Oxford. We thank S.~R.~Woodrow for her work on the trap design, T.~P.~Harty for contributions to the apparatus, and acknowledge the use of the University of Oxford Advanced Research Computing facility (doi:10.5281/zenodo.22558). The experiments benefitted from the use of the ARTIQ control system (doi:10.5281/zenodo.591804).

\section*{Author contributions}
\noindent CJB performed the numerical modelling. VMS and CJB designed and performed the experiments. KT built the ion trap and characterized the fast acousto-optic modulators. LJS and TGB built optical and control systems. VMS, CJB, AMS and DML wrote the manuscript, which all authors discussed.

\section*{Author information}
\noindent The authors declare no competing financial interests. Correspondence and requests for materials should be addressed to DML ({\tt d.lucas@physics.ox.ac.uk}).

\section*{Data availability}
\noindent The data that support the plots within this paper and other 
findings of this study are available from the corresponding author upon reasonable 
request.


\begin{thebibliography}{30}
\bibitem[{\citenamefont{footnote1}()}]{footnote1}
\bibinfo{note}{The minimum gate time given in ref.~\cite{Ballance2016}, 3.8\us, is the fwhm of the laser pulse; for fair comparison with the gates reported in the present work we quote the total gate time from the start of the rising edge to the end of the falling edge, which was 5.3\us.}
\end{thebibliography}
\end{document}